\documentclass[11pt,twoside]{article}
\usepackage{asp2006}
\usepackage{epsf}
\usepackage{psfig}
\usepackage{lscape}

\def\ltsim{\raise 2pt \hbox {$<$} \kern-1.1em \lower 4pt \hbox {$\sim$}}
\def\ltapprox{\raise 2pt \hbox {$<$} \kern-1.1em \lower 5pt \hbox {$\approx$}}
\def\gtsim{\raise 2pt \hbox {$>$} \kern-1.1em \lower 4pt \hbox {$\sim$}}
\def\gtapprox{\raise 2pt \hbox {$>$} \kern-1.1em \lower 5pt \hbox {$\approx$}}
\def\arcsec{$^{\prime\prime\,}$}

\def\rhalo{MRC~0116+111\,}

\def\com#1{$^\dagger$}

\begin{document}
\title{ Diffuse bubble-like radio-halo emission in MRC~0116+111: 
 Imprint of AGN feedback in a distant cluster of galaxies}   %%% Fill in title
\author{Joydeep Bagchi$^1$, Joe Jacob$^2$, Gopal-Krishna$^3$,
Nitin Wadnerkar$^4$, J. Belapure$^5$, Norbert Werner$^6$, A.C. Kumbharkhane$^4$}   %%% Fill in author names
\affil{$^1$IUCAA, Pune University Campus, Pune 411007, India \\
$^2$Newman College, Thodupuzha 685585, Kerala, India \\
$3$NCRA - TIFR, Pune University Campus, Pune 411007, India \\
$^4$School of Physical Sciences, S.R.T.M. University, Nanded 431606, India \\
$^5$Dept. of Physics, Pune University, Pune 411007, India \\ 
$^6$Kavli Institute for Particle Astrophysics and Cosmology, Stanford  
University, 382 Via Pueblo Mall, Stanford, CA 94305, USA}    %%% Fill in author affiliations

\begin{abstract} %%% Abstract to run on from here.
We report the discovery of a luminous,  mini radio halo of
$\sim$240 kpc dimension at the center of a distant cluster of galaxies at
redshift z = 0.131. Our optical and multi-wavelength GMRT and VLA
observations reveal a highly unusual structure
showing a twin bubble-like diffuse radio  halo surrounding a
cluster of bright elliptical galaxies; very similar to the large-scale
radio structure of M87, the dominant galaxy in Virgo cluster. 
It presents an  excellent 
opportunity to  understand the energetics and the dynamical  evolution of 
such radio jet inflated  plasma bubbles in the hot cluster atmosphere.

%	 The non-thermal energy content, pressure
%	  and enthalpy of these giant bubbles, which are buoyantly rising in the
%	   cluster atmosphere, and were probably inflated by radio-jets in the past,
%	    are estimated to be large enough to substantially affect any possible
%	     cooling-flow at the cluster center.

\end{abstract}

%%% MAIN BODY OF TEXT GOES HERE. CONSULT "INSTRUCTIONS FOR AUTHORS USING
%%% LATEX2E MARKUP", SECTIONS 2.3-2.6 FOR HELP WITH EQUATIONS, FIGURES,
%%% AND TABLES.
\section{Introduction}
Recent X-ray observations with {\it Chandra} and
{\it XMM-Newton} have revealed a surprising aspect of cooling flows in clusters;
they showed far less
cooling below X-ray temperatures than
expected, altering the previously accepted picture of
cooling flows \citep{Peterson06}. Unless gas is thermally supported, radiative
cooling  leads to a `cooling catastrophe', i.e.  
inexorable inflow of cold gas onto
the central galaxy. To prevent this, some  heating mechanism was required
to raise gas temperature above$\sim 2$ keV, supressing
the cooling flow. Although several such mechanisms were discussed, 
the most effective heating process is the energy injected into
the intra-cluster medium (ICM) by radio jets from AGNs
of central galaxies of  clusters and groups \citep{BT95,Churazov01} 
. Almost all cool-core clusters harbour powerful central 
AGNs \citep{burns90} which
%One of the canonical models of AGN posits accretion of gaseous
%medium as fuel for the nuclear
%black hole such that  AGN outflows are powered by
%gravitational binding energy released by the
%infalling matter \citep{Begelman84}. The observed strong association
%of AGN in the central galaxy and  the surrounding cooling-flow
%lends good support to
%this model. 
suggests that they are  fuelled by accretion of cooling gas
\citep{bk94,allen06}, with the flow rate
itself regulated by AGN-heating \citep{Churazov01}.
Many details of how this AGN-ICM feedback process works
are still far from clear. 
Radio jets from a central AGN would inflate two bubble-like lobes of non-thermal
plasma, filled with relativistic particles
and magnetic field, and thus become visible in radio observations \citep{GN73}. 
Such non-thermal bubbles are  visible in radio  and X-ray observations;  
such as those  
in clusters MS0735.6+7421, Hydra-A, and others, showing an
unusually large and energetic pair of radio emitting, X-ray dark cavities
(e.g., \citet{McNamara_et_al2000,McNamara05,McNamara07}). These  synchrotron 
plasma bubbles are 
resposible for the mechanical (PdV) work on the ICM for heating it, 
which is one of the suggested mechanism of AGN-ICM feedback. Therefore, 
observations of  bubble-like, diffuse
radio sources near  cluster centres can provide crucial data for  
understanding this important astrophysical process.

%\begin{figure}
%\plottwo{0116_GMRT_1280GHZSC5+IGO_R.eps}{spect_0116loglog_BW.ps,angle=90}
%\caption{{\itshape Left:\/} Fig 1a.
%{\itshape Right:\/} Fig 1b.}
%\end{figure}

%\begin{figure}
%\plottwo{0116_GMRT_610SC3+Bubble.eps}{m87_bubbles_label_color.eps}
%\caption{{\itshape Left:\/} Fig 1a.
%{\itshape Right:\/} Fig 1b.}
%\end{figure}
\begin{figure}
%\vbox{
\hbox{
\psfig{file=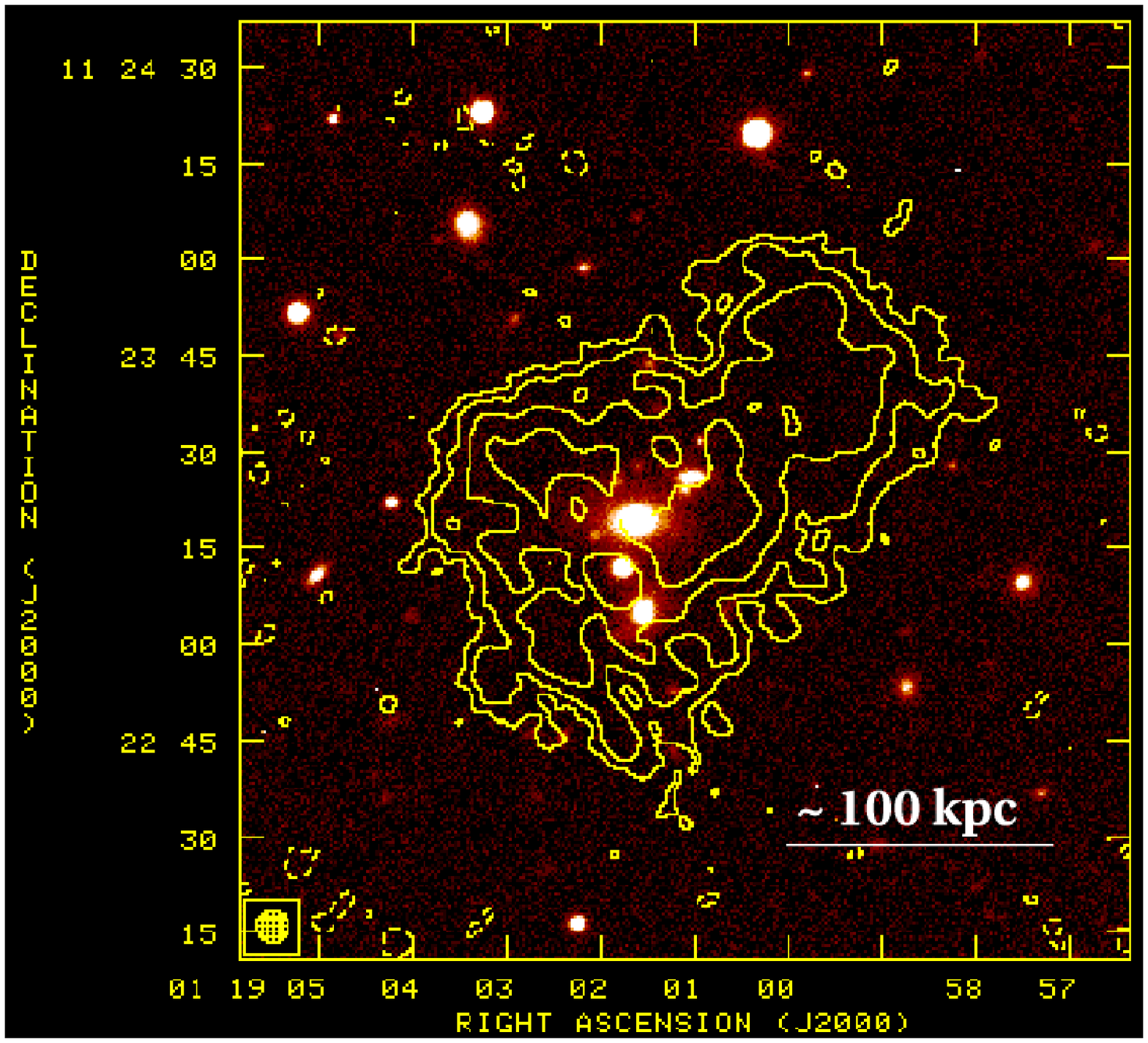,width=0.34\textwidth,angle=0.}
\psfig{file=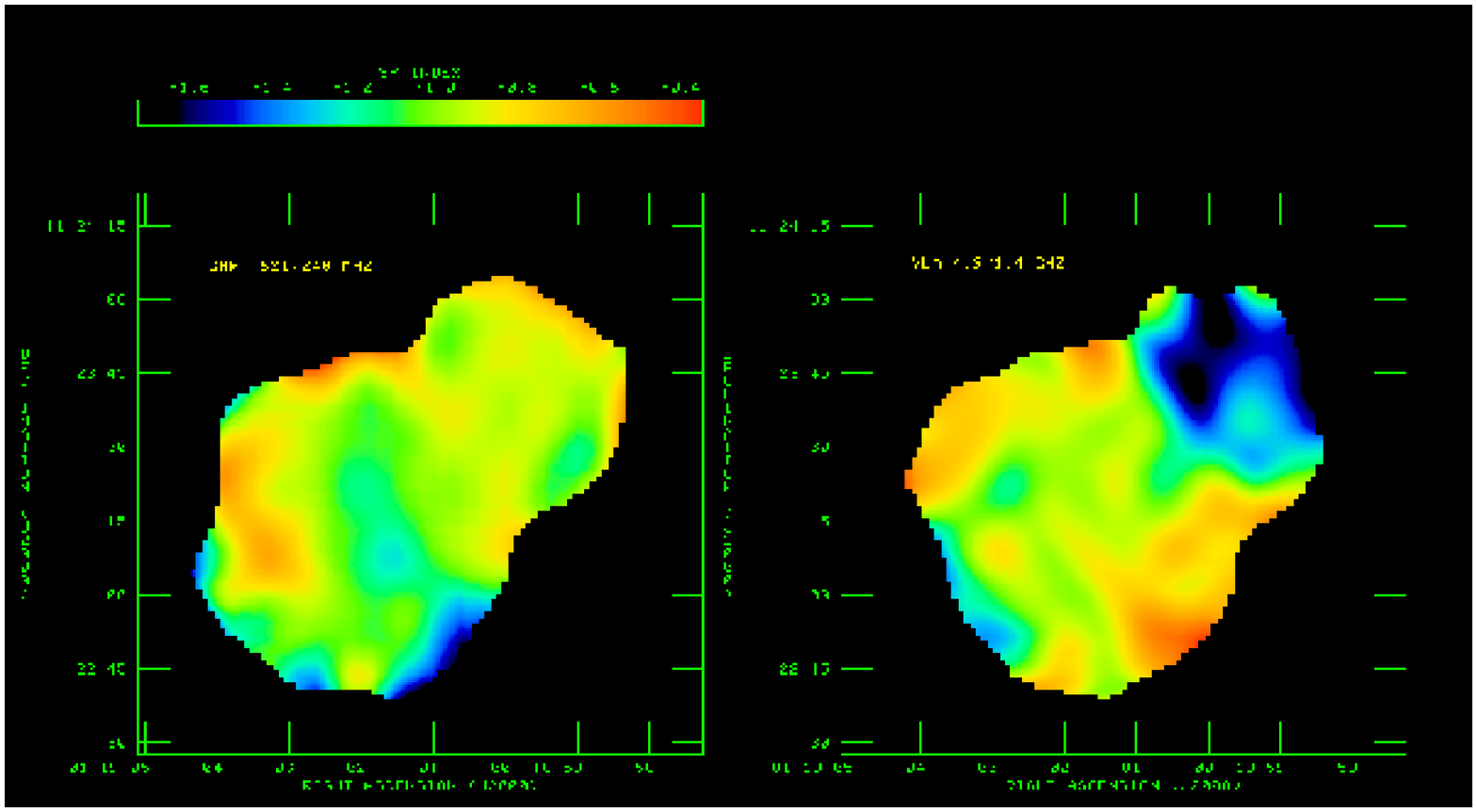,angle=-90.,width=0.64\textwidth}
}
%}
\caption{
{\bf left:} GMRT 1.28 GHz map of  MRC~0116+111 
(contour levels: $\pm$0.24, 0.48, 0.96, 2, 4, 8 mJy/beam, 
beam FWHM: 5\arcsec\, circular) overlayed on IGO 
R-band image. No AGN (radio core) 
is visible down to $\sim$ 1 mJy/beam
flux density limit, and no radio jets or lobes are detectable.
{\bf right:} Low and high frequency spectral index maps
of \rhalo  from combining the GMRT 
maps at 240 and 621 MHz (on left) and
VLA maps  at 1425 and 4860 MHz (on right). Both pairs of maps have the  
matched resolution of 12\arcsec\, (FWHM). Only pixels $\sim$3.5 times
above the noise level were included by giving  cut-offs
at values 4 mJy/beam, 0.5 mJy/beam, 0.27 mJy/beam and 0.15 mJy/beam
on 240, 621, 1425 and 4860 MHz maps respectively. 
}
\label{f1}
\end{figure}

\section{Optical \& radio observations: physical  picture of the source}
An early report on a diffuse radio source MRC~0116+111 matching the characteristics
of a mini-halo, was presented by some of us, based on  
VLA and GMRT observations \citep{IAU2002}. Recently,
a distant galaxy cluster was reported at the position of MRC~0116+111 
\citep{Lopes04}. Here we 
report higher sensitivity
GMRT observations made at 1.28 GHz,
621 MHz and 240 MHz  frequencies
using the  128 channel FX correlator. The earlier VLA observations made in 
C-band (DnC array) and L-band (CnB array) have been reanalysed. 
%For analysis standard routines in  AIPS were used. 
Optical broadband ({\it B,V,R,I})  CCD
imaging  observations were taken with IFOSC on the 2~mt telescope at the  
IUCAA  Girawali Observatory (IGO). For spectroscopy we used  the ESO  
3.6~mt New Technology Telescope (NTT) and EMMI (ESO Multi-Mode Instrument).
A low resolution slit-spectrum was taken with the grism-3 optics on the 
EMMI (December 1998) which gave a redshift $z=0.1316$ for the
brightest central cD galaxy and $z=0.1309$ for the second brightest
elliptical galaxy $\sim15^{''}$ south of cD \citep{IAU2002}. 

A radio-optical overlay of GMRT 1.28 GHz
radio map and optical CCD image of MRC~0116+111 (Fig.~\ref{f1}) shows a 
highly diffuse `halo' or `bubble' like structure which
bears close resemblance to  so-called `radio mini-halos' (RMH).
Mini-halos are $\sim$100 kpc scale, low surface brightness 
amorphous radio sources
with a steep spectral index ($\alpha$\ltsim\,-1, $S_{\nu}$ 
$\propto$ \, $\nu^{\alpha}$), which are rare objects and found around powerful AGNs 
at the center of cooling-core clusters \citep{Ferrari08}.
The 1.4 GHz radio luminosity of MRC~0116+111, 
$L_{\rm 1.4\,GHz}$ = $4.57 \times 10^{24}$ W/Hz, 
is quite large, comparable to the luminous radio-halos in Perseus 
and A2390 clusters \citep{pedlar90,Cassano08}.
Its  bolometric radio luminosity (over 10 MHz - 10 GHz range),
$L_{\rm radio}$ = $3.64 \times 10^{34}$ W, would place it amongst the most 
luminous radio halos known. 

%The high resolution
%1.28 GHz map (Fig.~\ref{f1}) and  the
%other radio maps do not show AGN-core and active radio-jets 
%feeding the 100-kpc scale twin plasma 
%lobes/bubbles of \rhalo. In this respect it 
%differs from  other known mini halos 
%\citep{Ferrari08,Cassano08,OEK,pedlar90} which have
%AGN cores and jets. Therefore, this suggests a physical
%picture in which the relativistic
%particles and magnetic fields  were probably seeded in the past by the
%nuclear  blackhole of the central cD galaxy via radio jets,
% as revealed by the
%pair of plasma bubbles centered around this galaxy (Fig.~\ref{f1}). Thus
%MRC~0116+111  is the unique example of a
%`fossil' cluster  mini-halo, 
%its twin extended lobes/bubbles of  radio plasma  
%not maintained by active jets from an AGN. 

\begin{figure}
\vbox{
\hbox{
\psfig{file=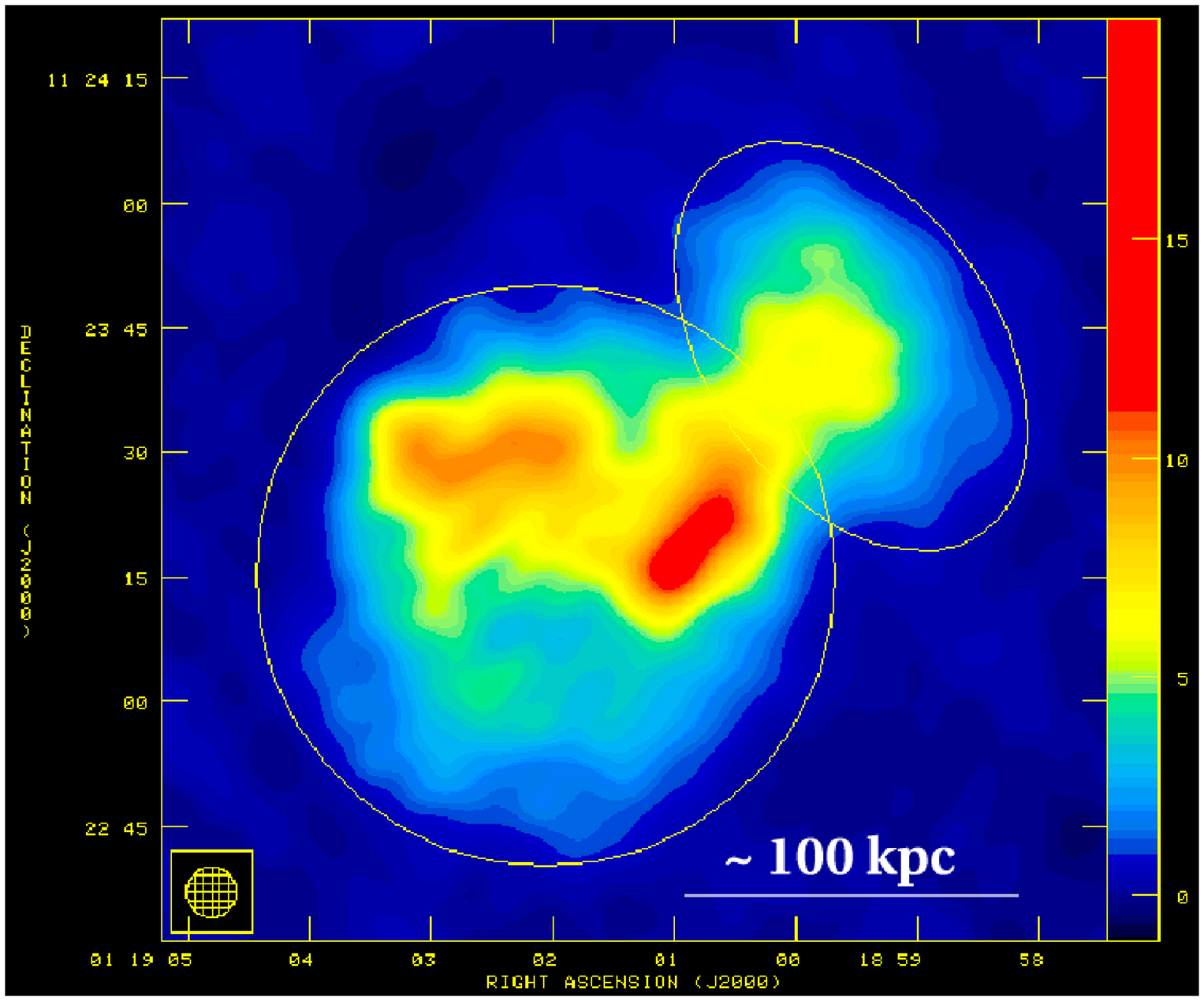,width=0.46\textwidth,angle=0.}
\psfig{file=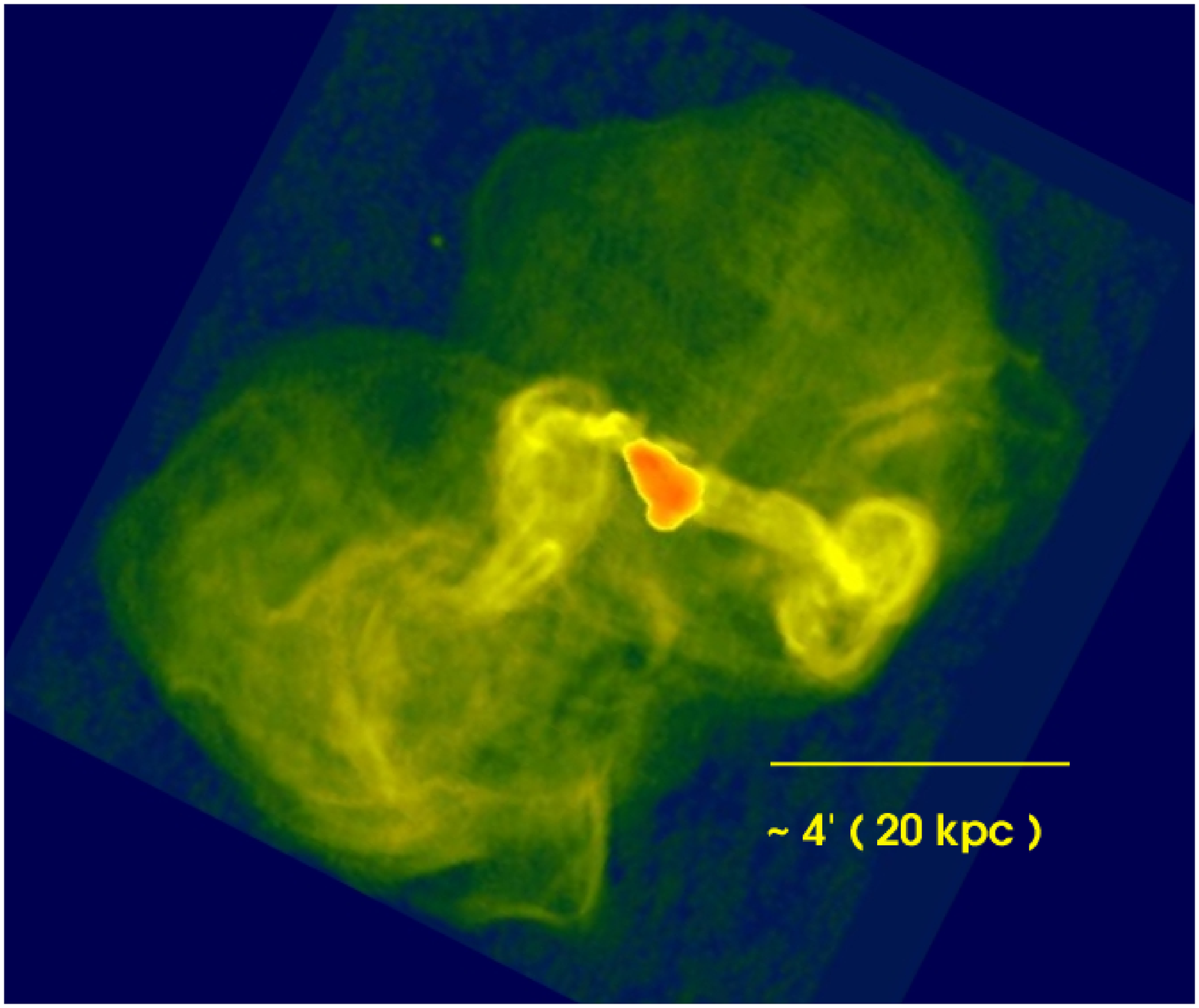,width=0.44\textwidth,angle=0.}
}
}
\caption{ {\bf left:}
GMRT 621 MHz grey-scale map  of MRC~0116+111
(beam: 6\arcsec\, circular) showing  diffuse radio-halo emission 
and twin bubble-like plasma structures which are delineated by 
dotted lines. {\bf right:} VLA 90~cm 
radio image of M87 in Virgo cluster (from
\citet{OEK}, reflected  and rotated by us). 
A remarkable morphological similarity can be seen
in both with twin bubble structures and similar internal outflow patterns.
}
\label{f2}
\end{figure}

Integrated spectrum of MRC~0116+111 shows a `break' near 400 MHz, after 
which it attains a power-law slope $\alpha = -1.35$ \citep{Bagchi09}, signifying
an ongoing particle injection, even though no radio core is detected 
down to 1 mJy level. Spectral index maps show that
between 240 and 621 MHz the spectral index distribution 
is fairly uniform ($\alpha_{mean} \approx -1$)  
with no strong steepening or gradients across the source, 
showing  a lack of strong radiative losses 
at these frequencies (Fig.~\ref{f1} center panel). 
However, the high frequency
spectral index map  between 1.4 and 4.8 GHz
shows a very different picture (Fig.~\ref{f1} right panel).
While the south-eastern bubble (Fig.~\ref{f2}) still has the same average
spectral index value around -1
($\alpha_{mean} = -1.06 \pm{0.15}$) - implying a straight synchrotron spectrum
between low and high radio frequencies - the north-west bubble, in contrast has 
developed an extremely steep spectrum ($\alpha_{mean} = -1.6 \pm{0.20}$), suggestive of
strong radiative energy loss in this part of the source.

Such a situation might  be explained if we assume that
the very steep spectrum north-western bubble
is buoyantly rising away and then detaching itself from a centrally located
mechanism of  energy injection, while this source or mechanism is
still active and possibly injecting  relativistic particles
into the flatter spectrum south-eastern bubble. One possibility is AGN activity,
though we fail to detect a radio core or jets (Fig.~\ref{f1}).
Based on a sample of 5-6 mini haloes, \citet{Gitti04} have reported
a positive correlation between the radio power of the mini-halo and
the cooling flow power. From this and because of their estimate of the
lifetimes of the radiating electrons falling considerably short of the
diffusion time scale,
% the similarity in the radio size to the cooling flow region,
they have inferred that in mini-haloes the cooling flow (through the
compressional work done on the ICM and the frozen-in magnetic field)
energizes the particle acceleration process through magneto-plasma
turbulence acting on relic electron population probably left behind by
the past episodes of AGN activity. 
%(also, Gitti et al. 2002).
An alternative proposal is that the radiating electrons in the mini-halos
are of secondary origin, created in proton-proton collisions 
\citep{Pfrommer_Ensslin04}.

%This injection mechanism can be an ongoing Fermi-type 
%re-acceleration
%of particles through shocks or turbulence resulting from 
%mergers \citep{Ferrari08}. 
%We can not exactly pinpoint how these giant plasma bubbles
%were created in this radio source. 
%As we discussed above,   
%possibly they were inflated  by radio-jets in a previous very
%energetic episode of AGN  activity. During the latter stages of evolution the
%dynamics of the remnants of the radio-jet  will be dominated by buoyancy forces
%and viscous drag (e.g., \citet{GN73,Churazov01}).

\section{Remarkable morphological similarity with M87 (Virgo A)}  

Figure.~\ref{f2} compares the GMRT 621 MHz radio map  with
the deep VLA 90~cm  image of  M87 \citep{OEK}, revealing striking morphological 
similarities. M87 shows  extensive diffuse outer structure extending
upto $\sim$40 kpc from the nucleus. Two collimated flows emerge from
the inner-jet region, one directed  north-eastward and the
other oppositely directed to the south-west (orientation as
shown in the rotated image, Fig.~\ref{f2}). The south-west flow ends 
in a well-defined pair of 
edge-brightened torus-like vortex rings.
The north-eastern flow develops a gradual but well-defined S-shaped
southward twist, starting only a few kiloparsecs beyond the inner lobes.
Finally, both flows are immersed in a pair of giant overlapping radio `bubbles', 
each extending about 40 kpc from the nucleus. After entering the outer lobes, 
the flows gradually disperse, filling the entire halo with radio-loud, 
filamented plasma.

We observe similar flow pattern in the central region of  MRC~0116+111: an 
edge brightened torus-like `mushroom' structure
about 40~kpc west of the center, analogous to the
peculiar `ear-shaped' vortex
structure of M87. Here the flow pattern sharply turns northward and
appear to be flowing into the smaller radio bubble to the north-west.
We observe an S-shaped flow pattern to the
east north-east of the center, which
further bends southward, and thus clearly resembles the  filamentary
structure visible in the southern bubble of M87 (Fig.~\ref{f2}). Both these 
radio sources are surrounded by a pair of radio emitting `bubble' like lobes.
The similarity of  large scale radio
outflow structures indicates that both sources might have originated in,
and their evolution goverened by, similar physical process and conditions
prevailing in the central regions of their host clusters.
Hydrodynamic simulation  of
\citet{Churazov01} suggests that
twin bubbles in M87 are buoyant bubbles of cosmic rays, inflated by jets
launched during an earlier nuclear active phase of the central galaxy,
which  rise through the cooling gas at roughly half the sound speed.
As shown above (section 2), detection
of a $\sim$100 kpc scale very steep-spectrum
radio bubble north-west of center of \rhalo strongly supports this model.
The flattened `mushroom' shape of this plasma bubble resembles a
rapidly rising vortex-ring (Fig.~\ref{f2}) into which an
intially spherical bubble will  naturally transform due to
viscosity and drag forces \citep{GN73,Churazov01}. Interestingly, 
at $\sim$240 kpc \rhalo is about
three times larger than M87 ($\sim$80 kpc; \citet{OEK}).
The bolometric radio luminosity of both is about the same,  
$L_{\rm radio}$ = $3.6 \times 10^{34}$ W, implying that average
volume emissivity of \rhalo is roughly factor 30 times smaller than that of M87.

\section{Summary \& future directions}
We have presented radio and optical observations of  MRC~0116+111 
showing that it is 
a $\sim$240 kpc scale radio mini-halo source 
%with no active nucleus or jets, 
in which huge radio bubbles
have been blown by a central AGN, and which are possibly
rising buoyantly in the hot atmosphere of a cluster. Such plasma bubbles
 are precursors of the giant X-ray dark cavities observed in several clusters.
A remarkable morphological similarity  with radio structure of M87
is found both in twin bubble-like structures and internal flow patterns,
suggesting a common origin. The lack of an X-ray image of the present 
cluster prohibits us from
quantifying the energy input rate by the AGN as well as the dynamics
of the mini-halo. This gap is in the process of being bridged.
Future X-ray imaging with {\it XMM-Newton} or {\it Chandra} will 
probe the intracluster medium, look for a cooling-core and search for
signs of radio plasma-ICM interaction: such as X-ray dark cavities, bright
filaments and  shock heating. 
Detection of giant cavities will allow  
calorimetry for estimating the energy injected into the cooling 
gas by the rising bubbles. It seems also feasible to search for 
non-thermal inverse-Compton X-ray emission of radio bubbles from  
lower energy electrons 
(with Lorentz factor $\gamma\, \sim \, $600 - 3000) up-scattering the CMB photons. 
 Combination of the observed 
non-thermal X-ray and radio fluxes  will determine the volume averaged 
magnetic field of intra-cluster medium in cluster core, which is currently unknown. 
%The upper limit on the X-ray emission from  MRC~0116+111 using ROSAT
%All Sky Survey (RASS) gives a lower limit on the volume averaged 
%magnetic field of $\sim$1.4 $\mu$G.

%\section*{}    %%% Unnumbered top level section head (remove "%" symbol)

%\subsection*{}   %%% Unnumbered second level section head (remove "%" symbol)

\acknowledgements %%% Text of acknowledgements runs on after this command.
We thank the operations team of the NCRA--TIFR GMRT observatory 
and IUCAA Girawali observatory. Norbert Werner was supported by the 
NASA through Chandra Postdoctoral Fellowship Award Number PF8-90056 issued  
by the Chandra X-ray observatory Center, which is operated by the  
Smithsonian Astrophysical Observatory for and on behalf of the  
National Aeronautics and Space Administration under contract NAS8-03060.

%%% THE BIBLIOGRAPHY
%%%
%%% CONSULT SECTION 3 OF "INSTRUCTIONS FOR AUTHORS" FOR HOW TO USE NATBIB.
%%% AUTHORS ARE ENCOURAGED TO USE EITHER THE "THEBIBLIOGRAPY" ENVIRONMENT
%%% BY UNCOMMENTING (DELETING THE "%" SYMBOL) THE COMMANDS BELOW, OR BY
%%% USING THE BIBTEX ENVIRONMENT. TO FIND OUT WHICH IS APPLICABLE TO YOUR
%%% CONTRIBUTION, CONSULT THE VOLUME EDITORS FOR YOUR PROCEEDINGS.
%%%

\end{document}